\documentclass[aps,nofootinbib,preprint,showpacs,superscriptaddress]{revtex4-1}
\date{\today}
\usepackage{graphicx,epsfig,latexsym,overpic,amssymb,color}
\usepackage{booktabs,longtable,supertabular}
\begin{document}
\title{Molecular structure of highly-excited resonant states in $^{24}$Mg
and the corresponding $^8$Be+$^{16}$O and $^{12}$C+$^{12}$C decays}
\author{Chen Xu}  
\affiliation{School of Physics, Peking University, Beijing 100871, China}
\author{Chong Qi}
\affiliation{Royal Institute of Technology (KTH), Alba Nova University Center, 
SE-10691 Stockholm, Sweden}
\author{R.J. Liotta}
\affiliation{Royal Institute of Technology (KTH), Alba Nova University Center,
SE-10691 Stockholm, Sweden}
\author{R. Wyss}
\affiliation{Royal Institute of Technology (KTH), Alba Nova University Center, 
SE-10691 Stockholm, Sweden}
\author{S.M. Wang}
\affiliation{School of Physics, and State Key Laboratory of Nuclear Physics 
and Technology, Peking University, Beijing 100871, China}
\author{F.R. Xu}
\affiliation{School of Physics, and State Key Laboratory of Nuclear Physics 
and Technology, Peking University, Beijing 100871, China}
\author{D.X. Jiang}
\affiliation{School of Physics, and State Key Laboratory of Nuclear Physics 
and Technology, Peking University, Beijing 100871, China}

\begin{abstract}
Exotic $^8$Be and $^{12}$C decays from high-lying resonances in 
$^{24}$Mg are analyzed in terms of a cluster model. 
The calculated quantities agree well with the corresponding experimental 
data.
It is found that the calculated decay widths are very sensitive to the
angular momentum carried by the outgoing cluster. It is shown that this
property makes cluster decay a powerful tool to determine the spin as well
as the molecular structures of the resonances.

\end{abstract}


\maketitle

\section{Introduction}

One of the most fundamental problems in theoretical nuclear physics has been
the determination of the degrees of freedom that govern the behavior of
the many-body system that is the nucleus. In the early 1930's it was suspected 
that the nucleus was like a polyatomic molecule or a liquid \cite{bra34}.
In this background the liquid drop model, with its collective degrees of
freedom, was introduced~\cite{boh39}. Also very early the
alpha-particle model of the nucleus was formulated~\cite{gam30,whe37}.
This model was successful to explain the structure of light $N=Z$ nuclei, e.g., 
$^{24}$Mg~\cite{mar86}. Even in heavy nuclei a model where the nucleus is
described as an alpha particle moving outside a frozen core~\cite{Bu90} 
has been applied successfully to explain nuclear properties (see, e.g., Ref.~\cite{Bu05} and references therein). 
This model was generalized to include exotic clusters as degrees 
of freedom and it could thus explain well cluster radioactivity \cite{Bu96}.   
But even before this, molecular-like states have been used as degrees of 
freedom since
they were first suggested \cite{bra34}. One expects that the cluster structures 
would be most noticeable in light $N=Z$ nuclei at excitation energies near 
the decay thresholds \cite{ikeda68,Freer07}. 

In this paper we will analyze cluster degrees of freedom in  $^{24}$Mg.
It has been found that this nucleus shows a rich variety of cluster structures. 
Already as early as 1960 resonant states observed in the $^{12}$C+$^{12}$C 
elastic scattering reaction were found to correspond to quasimolecular 
structures in the compound nucleus $^{24}$Mg~\cite{Bromley60,Vogt60}.
Thenceforth, many reaction experiments were performed where
different cluster-decay channels were observed, i.e.,
$\alpha$+$^{20}$Ne$^*$~\cite{Greenwood75,Ledoux84}, $^8$Be+$^{16}$O
\cite{Fletcher76,James78,Fulton89,Charissa94,Free98,Murgatroyd98,Free01},
$^{12}$C+$^{12}$C~\cite{Cosman80,Ledoux83,Costanzo91,Fulton94,Free98,Free01,Curtis95}. 
Also $\alpha$-chain structures have been suggested~\cite{Wuosmaa92}.

The question of whether the resonances observed in the $^{12}$C+$^{12}$C
scattering represent true cluster states in the $^{24}$Mg compound system 
or whether they simply reflect scattering states in the ion-ion potential
is still unsolved~\cite{Beck09}. Further, if correspondences indeed exist 
between the resonances and structures in $^{24}$Mg, one may ask which 
configuration would be responsible for the resonances. It was suggested that 
the resonant structures are associated with a strongly deformed secondary
minimum in the potential energy surface of $^{24}$Mg predicted by 
Nilsson-Strutinsky~\cite{Leander75,Aberg94} and Hartree-Fock~\cite{Flocard84}
calculations. The cranked cluster model also predicted several quasi-stable 
deformed cluster configurations at high deformations~\cite{mar86}.
The energy-spin plot of these resonant states indicates a large rotational
moment of inertia and hence a large deformation~\cite{Free97}. 
This was considered as an evidence for molecular structures.

Based on a $^{12}$C+$^{12}$C cluster model, Buck {\it et al.} has performed
coupled-channel calculations for $^{24}$Mg excited states in
excitation energy $\leq 20$ MeV, providing a good descriptions of 
spectroscopic properties including excitation energies, electromagnetic 
transitions and electron scattering~\cite{Buck90}. 
Similar spectroscopic studies but using the microscopic generator-coordinate 
method with $\alpha$+core configurations have also been carried 
out by Descouvemont and Baye~\cite{Descouvemont87,Descouvemont89}. 
Adopting the coupled-channel orthogonality condition model,
Kato {\it et al.} performed semi-microscopic calculations
for the band structures of the $^{12}$C+$^{12}$C and $^{16}$+2$\alpha$ 
molecular states~\cite{Ohkubo82,Kato89,Kato86}. However, a reliable 
calculation of charged-particle decay widths corresponding to 
resonant molecular states is still lacking. 
Yet, such calculation is necessary to get a further understanding of 
the observed resonant decays in $^{24}$Mg.

The width corresponding to the decay of a cluster from a mother nucleus is
given, in principle, by the classical Thomas expression~\cite{Tho54}, i.e., 
\begin{equation}\label{Tho}
\Gamma_l(R)=2{\cal P}_l(R) \frac{\hbar^2}{2\mu R} |{\cal F}_l(R)|^2,
\end{equation}
where $l$ is the angular momentum carried by the outgoing cluster,
${\cal P}$ is the penetration probability and
$\mu$ is the cluster-daughter reduced mass.
$R$ is a radius outside the surface of the daughter nucleus, where the
cluster is assumed to have been formed. At this point the nuclear
interaction acting upon the daughter nucleus and the cluster is negligible
and the outgoing channel
consists of two bodies, i.e., the daughter nucleus and the cluster, moving 
under the influence of the Coulomb and centrifugal forces only.
Inside this radius the nucleons forming the mother nucleus do not
necessarily  form any cluster.  The corresponding many-body wave function
should describe the clusterization as the relative distance $R$ approaches 
the nuclear surface.
At this point the wave function of the cluster already formed in the
internal region is matched with the corresponding outgoing
two-body wave function of the external region. The amplitude of the wave
function in the internal region at $R$ is the formation amplitude, i.e.,
\begin{equation}\label{foram}
{\cal F}_l(R)=\int d{\hat R} d\xi_d d\xi_c
[\Psi(\xi_d)\phi(\xi_c)Y_l(\hat R)]^*_{J_mM_m}
\Psi_m(\xi_d,\xi_c,\vec{R}),
\end{equation}
where $d$, $c$ and $m$ label the daughter, cluster and mother
nuclei, respectively. $\Psi$ are the intrinsic wave functions and 
$\xi$ the corresponding intrinsic coordinates. $\phi(\xi_c)$ is 
the intrinsic wave function of the cluster. The rest of the notation is 
standard. Notice that since at $R$ the internal and external wave functions
coincide one has $R {\cal P}_l(R)\propto 1/{\cal F}_l(R)$ and therefore 
the width is independent upon $R$ (for this and a simple derivation of
Eq.~(\ref{Tho}) see \cite{mag98}).
Notice also that the penetrability is determined by the Coulomb and
centrifugal forces only. Instead, the nuclear interaction is fundamental in the
formation of the cluster.

Among the observed exit channels of the $^{24}$Mg resonant states, 
decays into the $^8$Be$_{\rm gs}$+$^{16}$O$_{\rm gs}$ and
$^{12}$C$_{\rm gs}$+$^{12}$C$_{\rm gs}$ ones,
with all decay products in their ground
states, are of particular interest. Resonant states sampled through these
two channels have been expected to display quasimolecular
structures~\cite{Bromley60,Vogt60,Fulton89,Free98,Murgatroyd98,Free01,Beck09}.  
Experiments have shown that the $^8$Be and $^{12}$C decay channels emerge
mainly in the $^{24}$Mg excitation energy range of $\approx 23-34$
MeV~\cite{Fletcher76,James78,Fulton89,Costanzo91,Free98,Murgatroyd98,Free01}. 
The resonant widths of some of the states have been observed via
the $^{12}$C($^{12}$C,$^8$Be$_{\rm gs}$)$^{16}$O$_{\rm gs}$
reaction~\cite{Fletcher76,James78}.
An interesting question is if the same resonant states are
sampled in the two different decay channels~\cite{Free98}.

Besides the excitation energy and decay width, the spin is another important
quantum property of a resonant state. Usually, the spin is determined 
experimentally by analyzing angular correlation data. 
For most of the resonant states in $^{24}$Mg, however, spin assignments 
remain ambiguous due to statistical limitations of the data.  

We will here analyze the $^8$Be- and $^{12}$C-decays from the $^{24}$Mg 
resonant states. Our aim is to understand the processes leading to the
formation of the decaying resonances as well as to explore the structure
of the resonances in terms of molecular degrees of freedom.
In Section \ref{model} is the formalism to be used in the
applications presented in Section \ref{app}. A summary and conclusions
are in Section \ref{sum}.

\section{The model}
\label{model}
A proper (microscopic) evaluation of the formation amplitude
of the cluster is an extremely difficult undertaking.
It is for this reason that many effective models have been proposed.
These models take
somehow into account the formation probability through a number of free
parameters, which are adjusted to fit experimental data. With the parameters
thus determined, the effective models usually provide a good description
of the decay. Therefore effective models are very useful and extensively
applied, e.g., in Ref.~\cite{aru09}.
Among these models we will choose for our calculations the one described in 
Ref.~\cite{Bu90}, which has shown to provide the correct values of
cluster decay widths \cite{Bu05,Bu96}. This is an extreme cluster model, in 
which the mother nucleus is assumed to consist of the cluster moving around
the daughter nucleus (i.e., the core). In addition, it assumes that the decaying
state is quasibound and, therefore, that the mean-field potential generated
by the core can be viewed as a Woods-Saxon potential in which harmonic
oscillator conditions can be applied. Thus, a global principal quantum
number $G=2n+l$ is introduced, where $n$ is the number of nodes and $l$
the orbital angular momentum carried by the wave function. Applying the
Wildermuth condition (which is the harmonic oscillator condition for the
conservation of energy) the Pauli principle is partly taken into account.
As a result $G$ depends upon the number of nucleons in the daughter nucleus.

In this paper we will adopt the Woods-Saxon potential given by
\begin{equation}
\label{WSP}
v(r)=\frac{V_0}{1+e^{\frac{r-R}{a}}},
\end{equation}
where
\begin{equation}
V_0=-V_{00}(1\pm\kappa\frac{N_{\rm d}-Z_{\rm d}}{N_{\rm d}+Z_{\rm d}}),
\end{equation}
and the $+$ ($-$) sign corresponds to  proton (neutron) potentials.
The index $d$ indicates daughter quantities. 
%It is important to point out that we will not use any adjustable
%parameter. All the parameters in the theory will be taken from previous
%estimates. 
The Woods-Saxon parameters are from the Chepurnov parametrization of 
Ref.~\cite{Chepurnov}, i.e.,
\begin{equation}
\label{parameter}
\begin{array}{l}
V_{00}=53.3 ~{\rm MeV},\\
\kappa=0.63,\\
a=0.63 ~{\rm fm},\\
R=r_0{A_{\rm d}}^{1/3} ~{\rm fm},
\end{array}
\end{equation}
where $A_d=N_d+Z_d$, and $r_0=1.24$ fm (however, the parameter $r_0$ will be
adjusted to give a reasonable Coulomb barrier, see the discussion later).

We use a folded mean-field-type nuclear potential for the cluster,
which is constructed as follows~\cite{Xu06},
\begin{equation}
V_{\rm N}(r)=\lambda[N_{\rm c} v_{\rm n}(r)+Z_{\rm c} v_{\rm p}(r)],
\end{equation}
where $\lambda$ is the folding factor; $N_{\rm c}$ and $Z_{\rm c}$ are the
neutron and proton numbers of the cluster, respectively; $v_{\rm n}(r)$
and $v_{\rm p}(r)$ are the single neutron and proton potentials (excluding
the Coulomb potential), respectively, generated by the core, i.e., the
Woods-Saxon potential given by Eq.~(\ref{WSP}). 
Then the cluster potential is written as
\begin{equation}
V(r)=V_{\rm N}(r)+V_{\rm C}(r)+\frac{\hbar ^2}{2\mu r^2}l(l+1),
\end{equation}
where the Coulomb potential $V_{\rm C}(r)$ takes the usual form given 
in Ref.~\cite{Buck90} (taking the same radius for the Coulomb and nuclear
potentials).

We use the Bohr-Sommerfeld quantization condition~\cite{Bu90} in the 
determination of the parameter values, which is written as~\cite{Bu90}
\begin{equation}
\label{BS}
\int_0^{r_2} dr \sqrt{\frac{2\mu}{\hbar
^2}|Q_0-V(r)|}=(2n+1)\frac{\pi}{2}=(G-l+1)\frac{\pi}{2},
\end{equation}
where $Q_0$ is the decay $Q$ value for a given channel with the mother,
daughter and cluster all in their ground states; $r_2$ is
the turning point obtained by $V(r)=Q_0$. In the $Q_0$ case, we have $l=0$. 
For the $^{12}$C+$^{12}$C system,
the Coulomb barrier has been well measured experimentally~\cite{Kovar79}. 
In the present calculations, we adjust the radius parameter 
(i.e., $r_0$ in Eq.~(\ref{parameter}))
to fit the experimental Coulomb barrier by minimizing
$[(\frac{V_B^{\rm expt}-V_B^{\rm cal}}{V_B^{\rm expt}})^2+
(\frac{R_B^{\rm expt}-R_B^{\rm cal}}{R_B^{\rm expt}})^2]^{1/2}$ 
(here $V_B$ and $R_B$ are the height
and location of the Coulomb barrier, respectively). Therefore, the folding
factor $\lambda$ and the radius parameter $r_0$ can be determined by using the
Bohr-Sommerfeld condition and fitting the Coulomb barrier.

As seen, the Bohr-Sommerfeld condition involves the global quantum number $G$.
In our case the Wildermuth rule can be written as~\cite{Mo06} 
$G=\sum^{\rm A_c}_{i=1} g_i$, where ${\rm A_c}$ is
the nucleon number of the cluster and $g_i$ is the oscillator quantum number
corresponding to the nucleons in the cluster.
For the $^8$Be$_{\rm gs}$+$^{16}$O$_{\rm gs}$ structure in $^{24}$Mg, 
the $^8$Be cluster nucleons occupy the $1d_{5/2}$ orbits
above the $N,Z=8$ closed shells (i.e., the Fermi levels of the $^{16}$O core). 
These orbits have an oscillator quantum number $g_i=2$, leading to 
$G=16$ for the $^8$Be$_{\rm gs}$+$^{16}$O$_{\rm gs}$ configuration.
For the $^{12}$C$_{\rm gs}$+$^{12}$C$_{\rm gs}$ structure, 
two protons and two neutrons of the $^{12}$C cluster occupy the $1p_{1/2}$ 
orbits with $g_i=1$, and the other eight cluster nucleons fill the $1d_{5/2}$ 
orbits with $g_i=2$, which gives $G=20$. Notice that it is by considering 
the system as a Fermi gas that the Wildermuth rule takes into account the
Pauli principle. However, smaller $G$ values are possible if one 
considers the inner quantum of the cluster. For example, Buck {\it et al.}
take mainly $G=16$ as the starting number in the $^{12}$C+$^{12}$C
coupled-channel calculations~\cite{Buck90}. We will discuss calculations
at different $G$ values later.

The partial decay width is calculated by using the expression~\cite{Bu90}
\begin{equation}
\Gamma=P\times\Gamma_{\rm p},
\label{Gamma}
\end{equation}
and
\begin{equation}
\Gamma_{\rm p}=F\frac{\hbar^2}{4\mu}\exp\Big[-2\int^{r_3}_{r_2} k(r) dr \Big],
\label{Gamma_p}
\end{equation}
where $P$ is the preformation probability of the cluster being formed
in a state of the mother nucleus, and $\Gamma_{\rm p}$ is the width that 
corresponds to the penetration probability of the cluster through the 
potential barrier. The normalization factor $F$ is determined by
\begin{equation}
F\int^{r_2}_{r_1}\frac{dr}{2k(r)}=1,
\end{equation}
where $k(r)$ is the standard local wave number, i.e., 
\begin{equation}
k(r)=\sqrt{\frac{2\mu}{\hbar^2}|Q^*_l-V(r)|}.
\end{equation}
For the cluster decay from an excited state of 
the mother nucleus into the ground states of the decay products, we have
$Q_l^{*}=Q_0+E_J^{*}$ where $E_J^{*}$ is the excitation energy of the mother 
with the spin $J$ ($J=l$ for decay into spin-zero final states). 
$r_1$, $r_2$ and $r_3$ are the turning points obtained by $V(r)=Q^*_l$. 
The quantity $P$ is the probability that {\it within the model} the cluster
can be found in a given model configuration. It is the equivalent of the
spectroscopic factor in one-particle transfer reactions.
Thus, if there is only one configuration which is relevant,
then that configuration defines the decay channel and $P=1$. This was
the value used in, e.g., Refs.~\cite{Bu90,Bu96,Xu06}. 
This quantity has no relation with the cluster wave function on the nuclear
surface, except for the case of proton-decay, for which 
$P$ and the square of the formation amplitude
(\ref{foram}) are related. This is because the 
proton is indeed a ``cluster" in the sense of the
cluster model.  
But for real cluster (including the $\alpha$-particle) the preformation
factor $P$ is unrelated to
the formation amplitude  (\ref{foram}). Actually the values 
of the amplitude ${\cal F}$  
have been evaluated for alpha and heavier 
clusters by fitting experimental decay
widths. One thus obtained that ${\cal F}$ is of order unity for protons, 
as $P$ is, but of order $10^{-2}$ for alpha particles and of order
$10^{-5}$ for $^8{\rm Be}$ clusters~\cite{ir86}. The formation probability
is the square of this number.

\section{Calculations and discussions}
\label{app}
We take $G=20$ and 16 (i.e., ignoring the inner quantum number of the cluster)
for the calculations of the $^{12}$C$_{\rm gs}$+$^{12}$C$_{\rm gs}$ and 
$^8$Be$_{\rm gs}$+$^{16}$O$_{\rm gs}$ decays, respectively, as discussed above. 
Using the Bohr-Sommerfeld condition and fitting the Coulomb barrier,
with the experimental $Q$-value $Q_0=-13.93$ MeV for the 
$^{12}$C$_{\rm gs}$+$^{12}$C$_{\rm gs}$ channel, we obtained a folding
factor of $\lambda=0.500$ and an adjusted radius parameter of
$r_0=1.44$ fm for the $^{12}$C-cluster potential. This radius parameter
is slightly larger than the one of $r_0=1.24$ fm in the Chepurnov 
parametrization, which would indicate an equivalent inclusion of
the effect from the cluster size . 
For the $^8$Be$_{\rm gs}$+$^{16}$O$_{\rm gs}$ channel, no experimental
Coulomb barrier is available. We take the same radius parameter for this
channel, which should be reasonable. 
With $Q_0^{\rm expt}=-14.14$ MeV, we obtained $\lambda=0.509$ for the
$^8$Be-cluster potential. 

With all the quantities required by the formalism thus obtained we proceeded 
to analyze experimental data corresponding to resonances in $^{24}$Mg that have 
been measured at the high excitation energy-range  of $\approx 23-34$ MeV.
These resonances might be quasimolecular states since the emissions of 
$^8$Be- and $^{12}$C-clusters from the resonances have been observed 
in many experiments
\cite{Bromley60,Fletcher76,James78,Cosman80,Ledoux83,Fulton89,
Costanzo91,Fulton94,Charissa94,Free98,Murgatroyd98,Free01}.
The experimental widths of the resonances have been obtained via 
the measurement of the
$^{12}$C($^{12}$C, $^8$Be$_{\rm gs}$)$^{16}$O$_{\rm gs}$
channel~\cite{Fletcher76,James78}. The high resolution of $\approx 100$ keV in 
energy indicates that the observed widths would be the natural widths
of the resonances~\cite{Fletcher76,James78}. These can be considered
narrow resonances and, therefore, quasibound states. One of the main
assumptions of the cluster model is thus satisfied. 
To study the possibility that the resonances are indeed states of $^{24}$Mg 
we will evaluate the  $^8$Be- and $^{12}$C-decay widths emitted from all the
observed resonances. Those channels that provide values of the widths
that agree with experiment will give us a clue on the molecular structure of
the resonances as well as the corresponding spins. Regarding this last point
it is to be noticed that the decay channels 
$^8$Be$_{\rm gs}$+$^{16}$O$_{\rm gs}$ and
$^{12}$C$_{\rm gs}$+$^{12}$C$_{\rm gs}$ involve only spin-zero final
states and therefore $l=J$, where $l$ is the angular momentum of the
outgoing cluster and $J$ is the spin of the resonant state.
Moreover, in the $^{12}$C decay channel only
even angular momentum are carried by the outgoing $^{12}$C cluster.
This is due to the symmetric character of the
$^{12}$C + $^{12}$C partition in the entrance channel and that 
in the exit channel all states have zero spin.

With the folded $^8$Be-cluster potential obtained above, we calculated
the corresponding penetration width $\Gamma_{\rm p}$ (Eq.~(\ref{Gamma_p}))
for different spins. Since the centrifugal barrier (and therefore the 
penetrability) depends rather strongly upon $l$, this calculation may be 
able to determine accurately the spins of the observed resonant states. 
This feature is even more marked for the high-lying states, with energies
which may be above the Coulomb barrier.

\vspace{0.5cm}
\begin{longtable*}[htdp]{@{\extracolsep{32pt}}ccccccc}
\caption{\label{table1} Calculated penetration widths
$\Gamma_{\rm p}^{\rm cal}$ (Eq.~(\ref{Gamma_p})) corresponding to
the $^8$Be- and $^{12}$C-cluster decays from  $^{24}$Mg. 
The total width  $\Gamma^{\rm cal}$ is from Eq.~(\ref{Gamma}).
The experimental widths ($\Gamma^{\rm expt}$) are taken from
Ref.~\cite{James78}. The experimental spin assignments are taken from
Refs.~\cite{Cosman80,Free98,Free01}. The theoretical spins corresponding 
to the $^{12}$C channel must be even (see the text).}
\\[-0.5ex]
\hline \hline
$E_J^*$ (expt) & $\Gamma^{\rm expt}$ & $J^{\rm expt}$ &
$J^{\rm theo}$ & $\Gamma_{\rm p}^{\rm cal}$($^8$Be) & 
$\Gamma_{\rm p}^{\rm cal}$($^{12}$C) & 
$\Gamma^{\rm cal}$ \\
(MeV) & (keV) & & &(keV) & (keV) &(keV) \\
\hline
\endfirsthead
\multicolumn{3}{l}%
{\tablename\ \thetable{} -- continued} \\
\hline\hline
$E_J^*$ (expt) & $\Gamma^{\rm expt}$ & $J^{\rm expt}$ &
$J^{\rm theo}$ & $\Gamma_{\rm p}^{\rm cal}$($^8$Be) & 
$\Gamma_{\rm p}^{\rm cal}$($^{12}$C) & 
$\Gamma^{\rm cal}$ \\
(MeV) & (keV) & & &(keV) & (keV) &(keV) \\
\hline\\[-3.5ex]
\endhead
\\[-3.5ex]
\hline\\[-5.5ex]
 \multicolumn{7}{r}{{Continued...}} \\
\endfoot
\hline \hline
\endlastfoot
23.9  &  200  & (8) & 8 &  161 & 110  & 136 \\
24.2  & (200) & (8)   & 8 &  239 & 170  & 205 \\
24.4  & (400) & (8, 9)& 8 &  305 & 192  & 249 \\
      &       &       & 9 &  49 &      & 49    \\
24.6  &  300  & (8)   & 8 &  386 & 284  & 335 \\
      &       &       & 9 &  65 &      & 65    \\
24.9  & 300   & (8, 9) & 8 & 529 & 418  & 474 \\
      &       &       & 9 & 96  &      & 96    \\
25.1  &$<450$ & (8)   & 8 & 646 & 529 & 588\\
      &       &       & 9 & 124  &      & 124    \\
25.3  & 200   & (8)   & 8 & 759 & 628 & 694\\
      &       &       & 9 & 159  &      & 159    \\
25.8  & 500   &(9, 10) & 9 & 283  &      & 283    \\
      &       &       & 10& 40   & 37   &  39  \\
26.3  & 300   & (10)  & 9 & 474   &   &  474 \\
      &       &       &10& 74   & 68  &  71 \\
26.9  & 340   & (10)  & 10& 149  & 154  &  152 \\
27.3  & 300   & (10)  & 10& 230  & 234  & 232  \\
27.8  & 240   & (10)  & 10& 381  & 402  & 392  \\
28.3  & 340   & (10)  & 10& 600 & 654 & 627 \\
      &       &       & 11& 88  &      &  88    \\
29.3  & $\sim 700$ & (10,12) & 11 & 243 &  & 243   \\
      &       &       & 12& 28   & 38   & 33   \\
30.1  & $<400$ & (10, 12) & 11& 496 &      & 496    \\
      &       &       & 12& 64   & 91  & 78   \\
30.4  & $<400$&(12)& 12& 85   & 120  & 103  \\
31.1  & 320   & (12)  & 12& 165  & 244  & 205  \\
31.7  & 500   & (12)  & 12& 280  & 421  & 351  \\
32.7  & $\sim 500$& (12, 13) & 12 & 624  & 793 & 709 \\
      &       &       & 13& 76  &      &  76    \\
33.4  & 230   & (12, 13)  & 12& 963 &  & 963 \\
      &       &       & 13& 140  &      &  140    \\
\end{longtable*}

Proceeding in the same fashion, we also calculated the width 
corresponding to $^{12}$C-decay. The results of our calculations,
corresponding to both clusters, are presented in Table~\ref{table1}. 
One notices that the 
penetrability width $\Gamma_{\rm p}$ is approximately the same for both
channels for a given value of the angular momentum. This can be understood 
because at the high excitation energies of
the resonances the centrifugal barrier is dominant. 

An interesting question is whether the different choices of the $G$ number influence the 
calculated width. The global quantum number $G$ appears in the 
Bohr-Sommerfeld condition (i.e., Eq.(~\ref{BS})) which is used mainly in the
determination of the folding factor. Considering the inner quantum
of the cluster, however, smaller $G$ values are possible.
The $^8$Be cluster has an inner
quantum number of 4, and $^{12}$C itself has a number of 8, which gives 
a starting $G$ number of 12 for the two channels.
For highly-excited cluster states, larger $G$ numbers are also possible. 
As an example, Table~\ref{table2} lists calculated widths for the 23.9 MeV 
($J=8$) state (i.e., the first state of the Table~\ref{table1}) 
at different $G$ numbers. Note that the $G=20$ situation in the 
$^{12}$C channel corresponds to the $G=16$ case in the $^8$Be channel,
i.e., both ignore the inner quanta. We see that the calculated widths
as well as the obtained Coulomb barriers keep quite stable at different $G$
values. The experimental height and location of the $^{12}$C+$^{12}$C Coulomb
barrier are $V_B=5.8\pm0.3$ MeV and $R_B=6.5\pm0.4$ fm~\cite{Kovar79},
respectively, while Buck's model gave $V_B\approx 6.3$ MeV and
$R_B\approx 7.2$ fm~\cite{Buck90}.
Once the potential parameters are determined, the width calculation is
independent on the $G$ number because it does not appear in the formula of the
width (see Eq.~(\ref{Gamma_p})). Different $G$ numbers lead to the 
different bands of 
spectra~\cite{Buck90,Ohkubo82,Kato89,Kato86,Descouvemont87,Descouvemont89},
which would be discussed in our future work with the possible improvement
of the model. In the present work, we focus on the decay property of 
the resonances.

\begin{table}
\caption{\label{table2}Calculations at the different choices of the $G$ number.
Only the calculated width for the 23.9 MeV ($J=8$) resonance is 
presented as checking.}
\begin{ruledtabular}
\begin{tabular}{cccccc}
$G$ & $\lambda$ & $r_0$ (fm) & $V_B$ (MeV) & $R_B$ (fm) & 
$\Gamma_{\rm p}$ (keV) \\
\hline
\multicolumn{6}{c}{$^{12}$C+$^{12}$C}\\
16 & 0.320 & 1.57 & 6.53 & 7.24 & 106 \\
18 & 0.402 & 1.50 & 6.54 & 7.24 & 108 \\
20 & 0.500 & 1.44 & 6.54 & 7.23 & 110 \\
22 & 0.614 & 1.39 & 6.54 & 7.23 & 113 \\

\multicolumn{6}{c}{$^8$Be+$^{16}$O}\\
12 & 0.307 & 1.57 & 5.68 & 7.43 & 150 \\
14 & 0.399 & 1.50 & 5.68 & 7.42 & 153 \\
16 & 0.509 & 1.44 & 5.68 & 7.42  & 161 \\
18 & 0.636 & 1.39 & 5.67 & 7.44  & 176 \\
\end{tabular}
\end{ruledtabular}
\end{table}

To obtain the decay width (\ref{Gamma}) we have still to determine 
the values of the preformation factor $P$.
It was already mentioned that if there is only one channel which is open 
we assume $P\approx 1$. 
This assumption has been well tested within effective models for the cluster 
decays of the ground states in even-even heavy nuclei~\cite{Bu90,Ren04}. 
For highly-excited resonant states, other configurations with excited
cluster and/or excited core would exist, which decreases the preformation
factor of the interested channel,
and consequently reduces the corresponding decay width.
For the resonant states in the energy range of $\approx 23-34$ MeV, however,
all the experiments~\cite{Fletcher76,James78,Fulton89,Charissa94,Free98,
Murgatroyd98,Free01,Cosman80,Ledoux83,Costanzo91,Fulton94,Free98,Free01,
Curtis95} address that only the decay channels with all the decay products in
their ground states have been detected. This indicates that 
the $^{12}$C$_{\rm gs}$+$^{12}$C$_{\rm gs}$ and 
$^8$Be$_{\rm gs}$+$^{16}$O$_{\rm gs}$ configurations should be dominant
in the resonances of this energy range.

We can further analyze the effect on the width from other possible channels. 
The $^{12}$C$_{\rm gs}$+$^{12}$C($2_1^+$) 
channel with single $^{12}$C excitation to the $2_1^+$ (at 4.44 MeV) state 
should be next important compared with the ground-state channel.
This channel has a decay threshold 4.44 MeV higher
than the one of the $^{12}$C$_{\rm gs}$+$^{12}$C$_{\rm gs}$ channel,
but can lead to a possible spin decrease of 2 units for the cluster motion 
and then reduces the centrifugal barrier. As an example, 
we have calculated the width of the $^{12}$C$_{\rm gs}$+$^{12}$C($2_1^+$)
channel for the 23.9 MeV (J=8) state (correspondingly $l=6$), 
giving a width of only 0.3 keV for this excited channel.
Configurations with single and mutual $^{12}$C excitations to the $0_2^+$ 
state (i.e., the Hoyle state at 7.65 MeV) are interesting. 
The microscopic coupled-channel calculation predicted that
these configurations are of importance for resonances with excitation energy 
higher than 40 MeV~\cite{Ito02}. However, for the resonances in 
the energy of $20-34$ MeV, our calculations show
that decay widths for the channel with single $^{12}$C excitation to 
the Hoyle state are less than 1 keV, and much less for the one with 
double $^{12}$C excitations to the Hoyle state.
Therefore, for width calculations in the energy range studied, 
contributions from excited channels should be negligible. 

In cases where $^{24}$Mg decays by emitting $^8$Be as well as $^{12}$C 
one can assume $P(^8{\rm Be})+P(^{12}{\rm C})\approx 1$. But 
we have seen that the penetrability is about
the same for both cases. Therefore the system 
will be trapped within the barrier approximately the same time,
irrespective of whether it
is in the configuration $^8$Be$_{\rm gs}$+$^{16}$O$_{\rm gs}$ or in
$^{12}$C$_{\rm gs}$+$^{12}$C$_{\rm gs}$. 
In addition, experimental excitation functions indicate that
the decay into both channels proceeds through approximately the same 
probability \cite{Free97,Free98,Free01}. Therefore one may assume that
the preformation factor is $P\approx 0.5$ for $^8$Be as well as for $^{12}$C.
As a result, the  resonant widths acquire the form
$\Gamma\approx 0.5[\Gamma_{\rm p}(^8{\rm Be})+\Gamma_{\rm p}(^{12}{\rm C})]$.

One sees in Table~\ref{table1} that, for each resonance, there is a value of 
$\Gamma^{\rm cal}$ that agrees with the corresponding experimental data
within a factor of three. The quality of these numbers can best be
judged by noticing that in the simpler case of
$\alpha$-decay such agreement would be considered excellent \cite{del99}.
But the most important feature of Table~\ref{table1} is that it shows that 
cluster decay from highly excited resonances is a powerful tool to 
investigate the structure of the decaying resonances. This is because 
at high energies (above the Coulomb barrier) the decay width is 
practically only dependent upon the centrifugal barrier. 
As a result, the calculated widths are
very sensitive to the angular momenta carried by the decaying clusters.

Often the experimental spin assignment of the states in Table \ref{table1}
is uncertain due to the statistical limitations of angular correlation data 
\cite{Free98,Free01}. Our calculation confirms
the experimental assignments when doubtful, as e.g., for the state at 
$E^*_J=23.9$ MeV. In this context it is important to point out that 
our calculation definitely excludes a possible $J=6$ value of the spin of this
state, as was suggested in a previous experiment~\cite{James78}.
This is a general feature. The calculation helps to decide which 
spin is correct when there are several experimental possibilities. 
An example of this is the state at
$E^*_J=25.3$ MeV, where for $J$=9 the calculated decay
width agrees with the corresponding experimental value, while for $J$=8 the
difference between theory and experiment is more than a factor of three.

As a further test of the present model, we have also investigated 
the resonances in the energy range of $20-23$ MeV which were observed 
to have the $^{12}$C+$^{12}$C decay~\cite{Curtis95}. 
The experiment~\cite{Curtis95} was performed with 
a high resolution of 87 keV in energy, which is important to ensure that
the natural widths of resonances were measured. Table~\ref{table3} presents
the comparison between the calculated and experimental widths of 
the resonances. The experiment assigned the $J^\pi=4^+$ for the resonant 
states, but commented that $J^\pi=6^+$ remains tentative~\cite{Curtis95}. 
A later experiment pointed out that the $J^\pi=6^+$ is more possible for 
these resonances\cite{Free98}. See Table~\ref{table3}, it is found that 
observed widths increase in general with increasing energies, 
but a remarkable drop happens at 21.62 MeV. 
This would indicate a spin increase at this resonant state.
From the present calculations, we should be able to conclude that
the first two resonances of Table~\ref{table3} have $J^\pi=4^+$ and
all others have $J^\pi=6^+$.  

\begin{table}
\caption{\label{table3}Experimental and calculated decay widths for the
$^{12}$C+$^{12}$C resonances observed in the energy range of 
$20-23$ MeV~\cite{Curtis95}.}
\begin{ruledtabular}
\begin{tabular}{ccccc}
$E^*_J$ (expt) & $J^\pi_{\rm expt}$ & 
$\Gamma^{\rm expt}$ (keV) &
$\Gamma^{\rm cal}$ (keV) & $\Gamma^{\rm cal}$ (keV) \\
(MeV)  &    & & $J^\pi=4^+$ & $J^\pi=6^+$ \\
\hline
$20.77\pm0.02$ &   & $175\pm108$ & 159 & 11 \\
$21.18\pm0.02$ & $4^+$ & $219\pm69$ & 330 & 27 \\
$21.62\pm0.01$ & $4^+$ & $99\pm36$ & 589 & 64 \\
$21.82\pm0.01$ & $4^+$ & $97\pm29$ & 584 & 92 \\
$22.01\pm0.01$ & $4^+$ & $120\pm31$ &  & 128 \\
$22.26\pm0.02$ & $4^+$ & $123\pm53$ &  & 195 \\
$22.43\pm0.03$ & $4^+$ & $216\pm70$ &  & 253 \\
$22.99\pm0.03$ & ($6^+$) & $267\pm90$ &  & 535 \\
\end{tabular}
\end{ruledtabular}
\end{table}

One sees that for all states in Table~\ref{table1} the spins are relatively 
large. This can be understood considering that at excitation energies above 
the Coulomb barrier only the centrifugal force can trap the cluster 
inside the mother nucleus. In the framework of the shell model, when the 
cluster nucleons fill the $sd$ shell the maximum possible spin is 12 
for the $^8$Be+$^{16}$O configuration. This is an indication that for the
highest-lying states in Table~\ref{table1}, with spin $J=13$, 
the cluster nucleons occupy the $fp$ shell.

Another important feature is that the calculation allows one to get information
about the molecular structure of the resonances. This is otherwise a very
difficult undertaking. For instance, the resonance lying at 23.9 MeV
can be interpreted as a mixing of the molecular configurations 
$^8$Be+$^{16}$O and $^{12}$C+$^{12}$C. Instead, the state at 25.3 Mev is
predicted to have spin $J=9$ and consists entirely of the configuration
$^8$Be+$^{16}$O. The same procedure can be applied to all states in
Table~\ref{table1}.

Finally, it is interesting to notice that since for the $^{12}$C decay channel 
only even spins are possible, the number of molecular configurations is
limited. 

\section{Summary and Conclusion}
\label{sum}

In this paper we have analyzed the decay of $^8$Be and $^{12}$C clusters 
from $^{24}$Mg. The observed decaying resonances lie high in the spectrum,
at the energy range of $\approx 20-34$ MeV. Our aim was to probe whether
the clusters could be considered as elementary degrees
of freedom. This is reasonable since at such high energies the nuclear 
density is low and therefore the Pauli principle is not very effective in
hindering the cluster formation.
To evaluate the cluster-decay widths we applied
the cluster model of Ref.~\cite{Bu90}.
As a central field we chose the Woods-Saxon potential.
The results of the 
calculation are in agreement with the rather large amount of available
experimental data within a factor of three, as seen in Tables~\ref{table1} and
\ref{table3}.
Considering that these are complicated decay processes, such agreement
can be considered very good.

In many instances our calculation shows that the decay through the $^8$Be 
channel is as probable as the one proceeding through the $^{12}$C channel. 
We could thus conclude that, in terms of the cluster model, the decaying 
resonance corresponds to a mixing of the molecular states
$|^8{\rm Be} \otimes^{16}{\rm O}\rangle$ and 
$|^{12}{\rm C} \otimes^{12}{\rm C}\rangle$. This also
indicates that both decays occur from the same resonance and simultaneously,
a point that may be doubtful to an experimental observer. 

At the very high energies of the decaying resonances the centrifugal barrier
is practically the only one able to trap the cluster nucleons within the
mother nucleus.  Therefore the penetrabilities, and the resulting decay
widths, are very sensitive to the angular momentum carried by the outgoing
cluster. This peculiar feature allowed us to assign precisely the spins of the 
molecular states in $^{24}$Mg. This is specially
important in cases where there are several experimental possibilities for
the spin of a given resonance. We have thus found, e.g., that the
calculated width of a  state lying
at 33.4 MeV, which experimentally may have spin $J$ = 12 or 13, agrees with the
experimental value only if $J$=13.

In conclusion we have shown in this paper
that the analysis of cluster decay widths from
high lying resonances in light nuclei
is a powerful tool to determine the spins of the states as
well as their structures in terms of molecular degrees of freedom.

\section{Acknowledgments}

One of the authors (C.X.) would like to express his thanks to
Royal Institute of Technology (Alba Nova University Center)
for the hospitality extended to him during his stay there.
This work has been supported by the NSFC research grant (J0730316) for
undergraduates, the NSFC grants under Nos. 10735010 and 10975006,
the Chinese Major State Basic Research Development Program under 
Grant 2007CB815000, 
and the Swedish Research Council (VR).

\end{document}